\documentclass[apl,aps,twocolumn,amsmath,amssymb,superscriptaddress,floatfix]{revtex4}
\usepackage{graphicx}% Include figure files
\usepackage{dcolumn}% Align table columns on decimal point
\usepackage{bm}% bold math
\usepackage[sort&compress]{natbib}
\usepackage{xspace}
\usepackage{amsmath}
\usepackage{amssymb}
\usepackage{epsfig}

\newcommand{\nm}{ \,\text{nm}}

\newcommand{\mum}{ \,\mu\text{m}}

\begin{document}

\title{A circular dielectric grating for vertical extraction of single quantum dot emission}

\author{M. Davan\c co} \email{mdavanco@nist.gov}
\affiliation{Center for Nanoscale Science and Technology,
National Institute of Standards and Technology, Gaithersburg, MD
20899, USA}
\affiliation{Maryland NanoCenter, University of Maryland, College
Park, MD}
\author{M. T. Rakher}
\affiliation{Center for Nanoscale Science and Technology, National
Institute of Standards and Technology, Gaithersburg, MD 20899, USA}
\author{D. Schuh}
\affiliation{Institute for Experimental and Applied Physics, University of Regensburg, D-93053 Regensburg, Germany}
\author{A. Badolato}
\affiliation{Department of Physics and Astronomy, University of
Rochester, Rochester, New York 14627, USA}
\author{K. Srinivasan}
\affiliation{Center for Nanoscale Science and Technology, National
Institute of Standards and Technology, Gaithersburg, MD 20899, USA}

\date{\today}% It is always \today, today,

\begin{abstract}
\noindent We demonstrate a nanostructure composed of partially
etched annular trenches in a suspended GaAs membrane, designed for
efficient and moderately broadband ($\approx 5$ nm) emission
extraction from single InAs quantum dots. Simulations indicate that
a dipole embedded in the nanostructure center radiates upwards into
free space with a nearly Gaussian far field, allowing a collection
efficiency $>80~\%$ with a high numerical aperture (NA=0.7) optic,
and with $\approx12\times$ Purcell radiative rate enhancement.
Fabricated devices exhibit a $\approx10~\%$ photon collection
efficiency with a NA=0.42 objective, a $20\times$ improvement over
quantum dots in unpatterned GaAs. A fourfold exciton lifetime
reduction indicates moderate Purcell enhancement.
\end{abstract}

\pacs{78.55.-m, 78.67.Hc, 42.70.Qs, 42.60.Da} \maketitle

Efficient extraction of single photons emitted by individual
semiconductor epitaxial quantum dots (QDs) is a necessity for many
applications in spectroscopy and classical and quantum information
processing~\cite{ref:Shields_NPhot}. As epitaxially grown QDs are
embedded in semiconductor material, total internal reflection of the
emitted light at the semiconductor-air interface and radiation
divergence can typically lead to $<1~\%$ collection efficiencies
even with high numerical aperture (NA) optics. Photonic structures
such as micropillar cavities can provide both QD radiative rate
Purcell enhancement and a far-field radiation pattern that can be
effectively collected~\cite{ref:Strauf_NPhot}, but require precise
spectral tuning of the cavity resonance to the QD emission line. In
contrast, vertically-oriented etched nanowires~\cite{ref:Claudon}
are spectrally broadband structures that have recently been shown to
provide large free space collection efficiencies, albeit without
Purcell enhancement and with an involved fabrication process.
Broadband operation not only relaxes the spectral alignment
requirement, which may impose strict constraints in fabrication
tolerances, but is also a necessity in spectroscopic applications in
which simultaneous detection of various spectrally separate
transitions is desired. Here, we present an approach for efficient
free space extraction of QD emission using a suspended circular
grating. This structure requires a simple nanofabrication procedure,
and supports a relatively broad (few nanometer) optical resonance
with a directional, nearly Gaussian far-field, which allows
efficient free space photon collection. Simulations predict a
collection efficiency of $\approx53$~\% ($80$~\%) into a NA=0.42
(0.7) optic. In fabricated devices, we report a $\approx10~\%$
single QD photoluminescence (PL) collection efficiency into a
NA=0.42 objective, a $\approx20\times$ improvement compared to QDs
in unpatterned bulk GaAs. A fourfold reduction in QD lifetime is
also observed, indicating moderate radiative rate enhancement.

\begin{figure}[]
\centerline{\includegraphics[width=8.5cm]{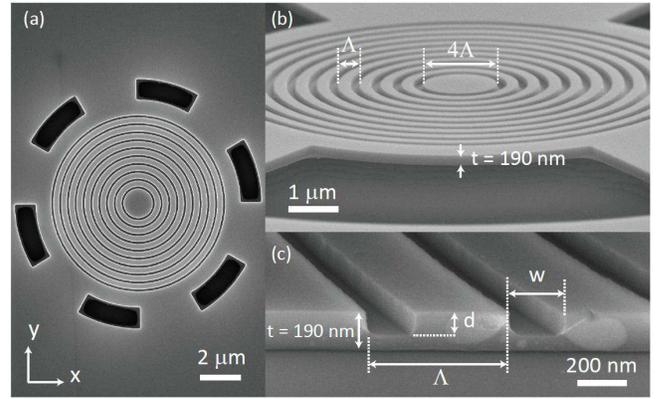}} \caption{(a)
Top, (b) angled, and (c) cross-sectional SEM images of suspended
circular dielectric grating structure.} \label{FIG1}
\end{figure}

Our nanostructure (Fig.~\ref{FIG1}) consists of a circular
dielectric grating with radial period $\Lambda$ that surrounds a
central circular region of radius $2\Lambda$, produced on a
suspended GaAs slab of thickness $t=190$~nm. The GaAs slab supports
single TE and TM polarized modes (electric or magnetic field
parallel to the slab, respectively). The grating is composed of ten
partially etched circular trenches of width $w$ and depth $d$, with
$t/2<d<t$. Quantum dots are grown at half the GaAs slab thickness
($z=0$), and located randomly in the $xy$ plane. This 'bullseye'
geometry favors extraction of emission from QDs in the central
circular region. It is based on (linear) high-contrast second-order
Bragg gratings recently introduced~\cite{ref:Baets1} for light
extraction from planar waveguides. While similar circular geometries
have been employed for enhanced light extraction from light emitting
diodes~\cite{ref:su_APL_033105}, and for demonstrating annular Bragg
lasers~\cite{ref:Green}, here we show an application in QD single
photon extraction.

\begin{figure}[t]
\centerline{\includegraphics[width=8.5cm]{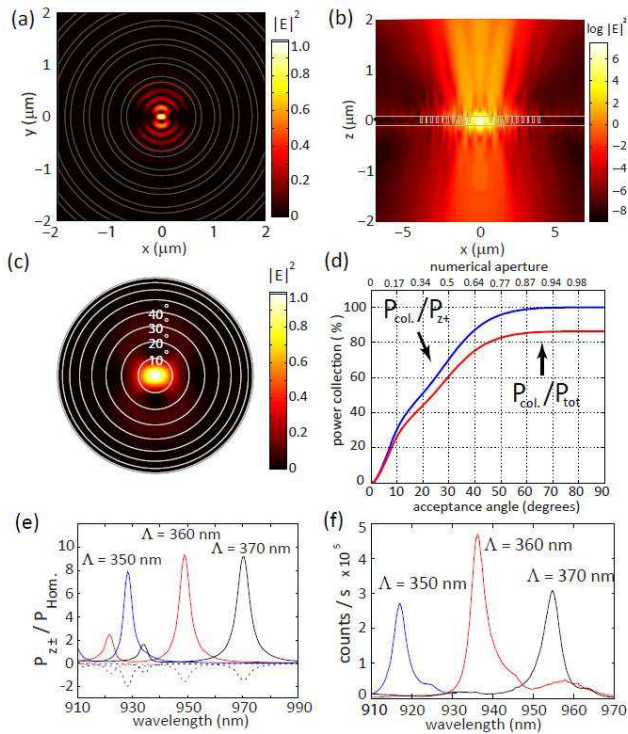}}
\caption{Electric field intensity in the (a) $xy$ and (b) $xz$
planes (log scale). (c) Far-field polar plot for the cavity mode
with $\Lambda=360$~\nm. (d) Collected power ($P_\text{col}$) as a
function of varying NA, normalized by the upwards ($P_{z+}$) and
total ($P_\text{tot}$) emitted powers. (e) Calculated vertically
extracted power as a function of wavelength, normalized to the
homogeneous medium electric dipole power $P_{\text{Hom}}$ for
$d=0.70t$. Continuous lines: upwards ($+z$) extraction; dotted:
downwards ($-z$). (f) Experimental PL spectra for high QD density
devices.} \label{FIG2}
\end{figure}

The design process consisted of a series of finite difference time
domain simulations that maximized vertical light extraction near the
expected QD s-shell emission ($\lambda_{QD}\approx940$~nm), by
varying $\Lambda$, $t$, and $w$. The structures were excited with a
horizontally oriented electric dipole at the bullseye center
$(x=0,y=0)$, representing an optimally placed QD. Total radiated
power, steady-state upwards emission, and electromagnetic fields
were then recorded at several wavelengths. The grating period
$\Lambda$ was initially chosen to satisfy the second-order Bragg
condition, $\Lambda=\lambda_{QD}/n_{TE}$, to allow for efficient
vertical light extraction ($n_{TE}$ is the GaAs slab TE mode
effective index). The dipole orientation was assumed to be aligned
along the $xy$ plane, exciting only TE slab waves. Starting values
for trench width and depth were $w=100$~nm and $d=0.5t$, deemed to
be easily fabricated. Vertical light scattering at the gratings is
partial, so that second-order Bragg reflections towards the center
lead to vertically leaky cavity resonances as shown in
Figs.~\ref{FIG2}(a) and (b). The large index contrast at the
trenches leads to strong reflections and out-of-slab-plane
scattering at the semiconductor-air interfaces, evident in the
strong field concentration at the bullseye center in
Fig.~\ref{FIG2}(a) and the fast field decay within the first couple
of trenches from the center (Fig.~\ref{FIG2}(b)). Large differences
in propagation constants in the semiconductor and air produce
significant resonance spectral shifts with small variations in
trench width. Trench depth ($d$) has a strong influence on the
quality factor ($Q$) and vertical light extraction, as incomplete
spatial overlap between a trench and an incident slab-bound wave
leads to both coupling to radiating waves and lower modal
reflectivity. Preferential upwards vertical extraction results from
the grating asymmetry, and is optimized through the trench
depth~\cite{be_apl_note}. We note that in addition to the mode shown
in Fig.~\ref{FIG1}, the cavity supports resonances which can be
excited by dipoles offset from the bullseye center. Coupling to
these resonances can lead to modified spontaneous emission rates and
collection efficiencies~\cite{be_apl_note}.

Figure~\ref{FIG2}(e) shows simulated, upwards (continuous) and
downwards (dotted) vertically extracted power as a function of
wavelength for structures with $\Lambda=350$ nm, 360~nm, and 370~nm,
$w=110$~nm, and $d\approx0.70t$. All curves are normalized to the
homogeneous medium electric dipole power, $P_{\text{Hom}}$. Trench
parameters reflect a trade-off in cavity $Q$ and vertical light
extraction, as discussed above. It is apparent that for each
$\Lambda$, an $\approx5$ nm wide resonance exists, with preferential
upwards ($+z$) light extraction. The upwards extracted power is
$\approx10\times P_{\text{Hom}}$, an indication of Purcell radiation
rate enhancement due to the cavity~\cite{ref:Vuckovic1}. Indeed, for
the $\Lambda=360$ nm structure, on which we now focus, the
enhancement $F_p$ at the maximum extraction wavelength ($\lambda_c =
948.9$ nm) is $F_p=P_\text{tot}/P_\text{Hom}=11.0$, where
$P_\text{tot}$ is the total radiated power in all directions. This
resonance has $Q=200$, and its effective mode volume, calculated
from the field distribution, is $V_\text{eff}=1.29(\lambda_c/n)^3$
($n$ is the GaAs refractive index)~\cite{be_apl_note}. The value for
$F_p$ predicted by $Q$ and $V_{\text{eff}}$ is $\approx$11.8, and is
consistent with the value determined above by the dipole radiation
simulations. Note that, given the modal field distribution in
Fig.~\ref{FIG2}(a), the modified dipole emission rate depends
strongly on its spatial location, being maximal at the bullseye
center.

The steady-state fields at a surface just above the GaAs slab were
used to calculate the far-field pattern in Fig.~\ref{FIG2}(c), which
shows that the emission is nearly Gaussian with a small divergence
angle. To better quantify this, we calculate the power
$P_\text{col}$ collected by an optic of varying NA.
Figure~\ref{FIG2}(d) shows the fractions of the upwards emitted
($P_{z+}$) and total ($P_{tot}$) powers collected as a function of
the collection optic acceptance angle. For NA=$0.42$ ($24.8^\circ$
acceptance angle), $\approx60~\%$ of the upwards emitted power (or
$\approx53~\%$ of the total emission) can be collected. For
NA$>0.7$, or an acceptance angle $>44.4^\circ$, collection superior
to 80~\% of the total emission can be achieved. We note that our
suspended grating approach limits radiation into the substrate
without the need to oxidize the AlGaAs, bond the grating to a low
index layer\cite{ref:Green}, or utilize a deeply etched
geometry~\cite{ref:Strauf_NPhot,ref:Claudon}.

Gratings were fabricated in a $t=190$~nm GaAs layer containing a
single layer of InAs QDs (density gradient from $>100~\mum^{-2}$ to
0 $\mum^{-2}$ along the $(01\bar{1})$ direction) on top of a
1~$\mum$ thick $\text{Al}_{0.6}\text{Ga}_{0.4}\text{As}$ sacrificial
layer~\cite{be_apl_note}. Fabrication steps included electron-beam
lithography, plasma dry etching, and wet chemical etching. The
plasma dry etch was optimized so that the GaAs would be partially
etched to a desired depth in the grating region
(Fig.~\ref{FIG1}(b)), and fully etched over the curved rectangles
just outside the grating region (Fig.~\ref{FIG1}(a)), which were
used in the wet etching step to undercut and suspend the device.

\begin{figure}[]
\centerline{\includegraphics[width=8.5cm]{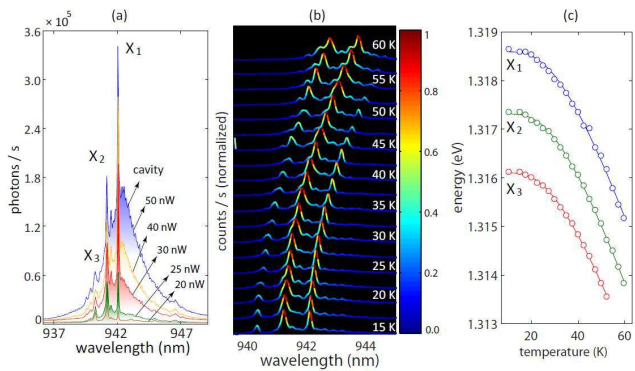}} \caption{(a)
PL spectrum from a low QD density $\Lambda=360$~nm device, for
various pump powers. (b) Temperature evolution of spectrum in (a)
(25 nW pump). (c) Temperature evolution of excitonic energies.
Continuous lines are fits.} \label{FIG3}
\end{figure}

Testing was done in a liquid He flow cryostat at $\approx8$ K.
Figure~\ref{FIG2}(f) shows PL spectra of three devices with a high
QD density and $\Lambda=$350\~nm, 360~nm, and 370~nm and $d/t>0.7$,
for pulsed pumping at a 780~nm wavelength (above the GaAs bandgap).
The spectra closely resemble the theoretical curves of
Fig.~\ref{FIG2}(e), with three, $\approx 5$ nm wide peaks spaced by
$\approx20$~nm. Deviations are likely due to differences in geometry
and refractive index between simulated and fabricated structures.
These results validated our simulations, and served to calibrate the
fabrication process. Figure~\ref{FIG3}(a) shows PL spectra at
various pump powers for a device with $\Lambda=360$~nm, now produced
in a low QD density region of the sample. Three isolated exciton
lines are observed on top of a broad background near 942~nm. The
sharp lines red-shift with increasing temperature
(Fig.~\ref{FIG3}(b)) with a dependence that can be fit to a model
that predicts a red-shift of the InAs bandgap (Fig.~\ref{FIG3}(c))
~\cite{ref:ortner_PhysRevB.72.085328,ref:kroner_temperature_2009,be_apl_note}.
In contrast, the broad background observed in Fig.~\ref{FIG3}(a)
shifts more slowly with temperature, and likely originates from
out-coupling of broad QD multiexcitonic emission via the leaky
cavity mode~\cite{ref:Winger2}. This is reinforced by the
observation, in Fig.~\ref{FIG3}(b), that the sharp QD lines are
maximized in the wavelength range
$940~\text{nm}<\lambda<942~\text{nm}$, when aligned to the broad
cavity peak, and decrease when driven away from it. The slower
cavity mode shift with temperature corresponds to a shift in
refractive index~\cite{ref:Badolato}.

\begin{figure}[h!]
\centerline{\includegraphics[width=8.5cm]{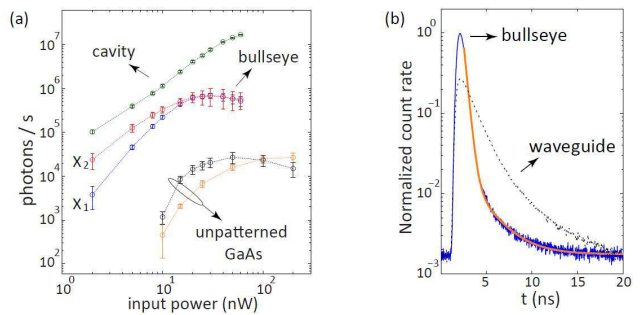}} \caption{(a)
PL as a function of pump power for $\text{X}_1$, $\text{X}_2$ and
cavity emission from Fig.~\ref{FIG3}(a), and two QDs in unpatterned
GaAs. Error bars are 95\% fit confidence intervals. (b) Solid:
$\text{X}_1$ lifetime trace with fit. Dotted: lifetime trace for QD
embedded in a suspended GaAs waveguide.} \label{FIG4}
\end{figure}

Figure~\ref{FIG4}(a) shows the detected PL as a function of average
pump power for the excitonic lines $\text{X}_1$ and $\text{X}_2$ and
the cavity mode emission from Fig.~\ref{FIG3}(a). While $\text{X}_1$
and $\text{X}_2$ saturate at $\approx20$~nW, the cavity emission
increases past this level. This further supports our assignments of
QD transitions and cavity mode in the Fig.~\ref{FIG3}(a) spectra.
Saturated photon rates (collected with a NA=0.42 objective) from
$\text{X}_1$ and $\text{X}_2$ were at least 20 times higher than
from typical QDs embedded in unpatterned GaAs, as shown in
Fig.~\ref{FIG4}(a). Assuming 100~\% QD quantum efficiency, we
estimate a collection efficiency of $\approx10~\%$ is achieved with
the bullseye pattern~\cite{be_apl_note}. A lifetime measurement of
$\text{X}_1$ after a $\approx300$ pm bandpass filter
(Fig.~\ref{FIG4}(b)) exhibits a multi-exponential decay with a fast
lifetime of $\approx360~\text{ps}$, limited by the $\approx600$~ps
timing jitter of the detectors. For comparison, the lifetime of a
single QD inside of a suspended GaAs
waveguide~\cite{ref:davanco_wgAPL} (dotted in Fig.~\ref{FIG4}(b)),
for which no radiative rate modification is expected, was
$\approx1.5$ ns. This suggests $F_p>4$. Note that since the pump in
Fig.~\ref{FIG4}(a) is pulsed with a 20 ns repetition period,
significantly longer than the lifetime, the increase in detected
counts relative to unpatterned GaAs is solely due to enhanced photon
extraction and collection into the objective.

Improved photon extraction efficiency can potentially be achieved
with a higher NA collection optic (50\% increase for NA=0.7) and
fabrication control~\cite{be_apl_note}, while deterministic QD
spatial alignment ~\cite{ref:Hennessy3,ref:Thon} can enhance both
the efficiency and Purcell factor. Although single-photon emission
from the bullseye is accompanied by undesirable cavity emission, a
few devices exhibited considerably less cavity mode feeding, albeit
with lesser extraction efficiencies. Since enhanced extraction
efficiency is due to the directional far-field pattern, a trade-off
may be achieved between Purcell enhancement and cavity feeding for
reduced $Q$. It is also likely that quasi-resonant QD pumping will
lead to reduced cavity feeding~\cite{ref:Ates_PRL}. These
possibilities are under investigation.

In summary, we developed a nanophotonic circular grating that
provides $\approx 10~\%$ free space collection efficiencies for
single InAs QD photons within a wavelength range of $\approx5$~nm.
Lifetime reduction of at least a factor of four is achieved, which,
taken together with the enhanced collection efficiency, indicates
Purcell radiative rate enhancement. This structure allows for
efficient and broadband spectroscopy of single QDs, and has
potential for use as a bright single-photon source.

The authors acknowledge the help of Robert Hoyt. This work has been
partly supported by the NIST-CNST/UMD-NanoCenter Cooperative
Agreement.

\bibliographystyle{apsrev}
%\bibliography{KS_bib_12_20_2010}

\newpage
%\cleardoublepage

\onecolumngrid \bigskip
\appendix
\setcounter{figure}{0}
\begin{center} {{\bf \large SUPPORTING
INFORMATION}}\end{center}
\subsection{Design simulations} 
\begin{figure}[h]
\centerline{\includegraphics[width=14cm,trim=0 0 0 0]{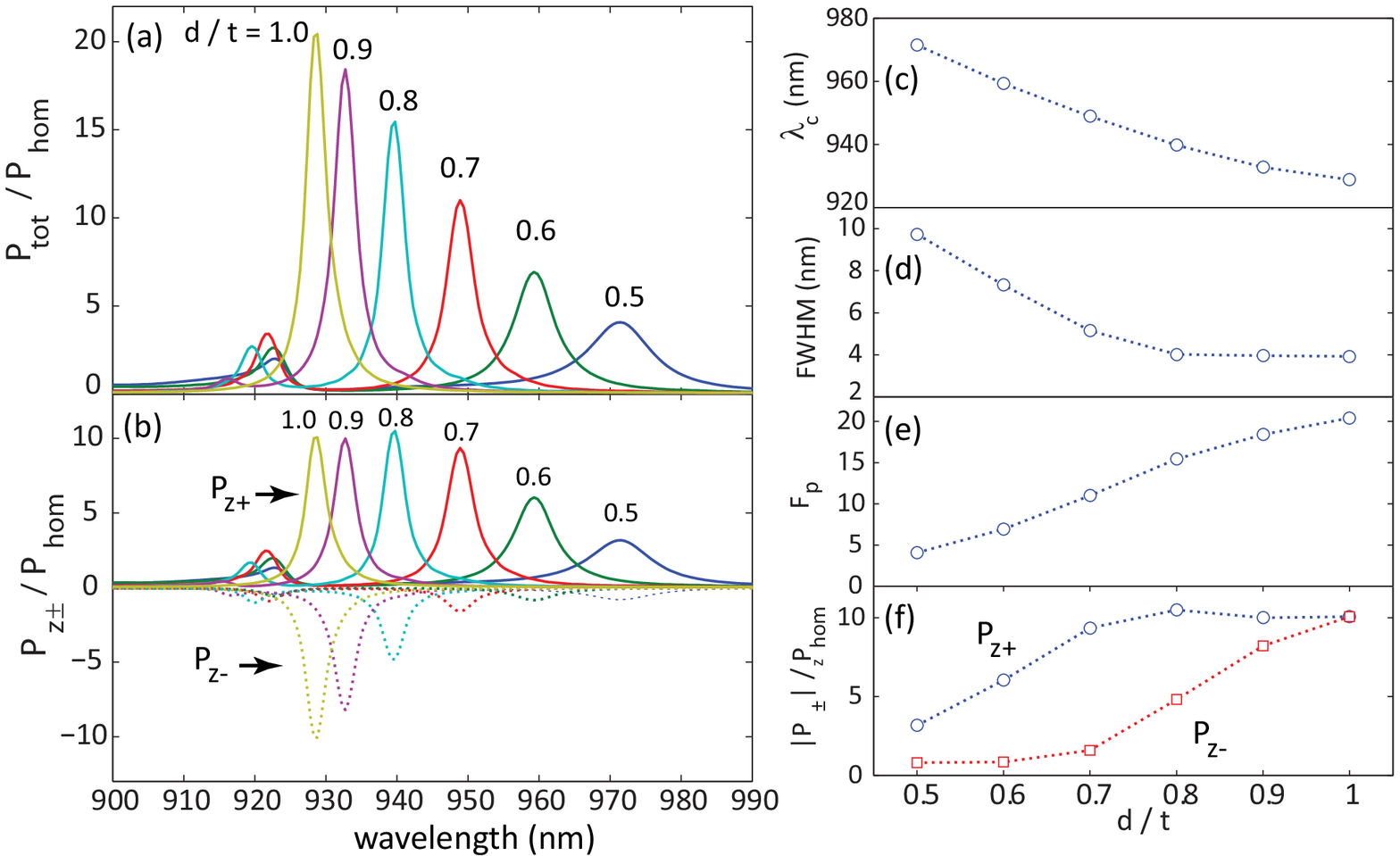}}
\caption{(a) Total emitted power $P_\text{tot}$ as function of
wavelength for various trench depths. (b) Vertically emitted power
$\pm z$ direction (continuous:$P_{z+}$, upwards; dotted:$P_{z-}$,
downwards), as a function of wavelength. (c) Central wavelength, (d)
full width at half maximum, (e) Purcell enhancement factor and (f)
maximum upwards ($P_{z+}$) and downwards ($P_{z-}$) emitted power as
functions of trench etch depth $d$. $P_\text{hom}$ is the
homogeneous space electric dipole emitted power and $t$ is the GaAs
slab thickness.} \label{FIG1X}
\end{figure}

The following simulation results
illustrate the effects of varying trench depths on emission
properties of the circular dielectric grating. Figure~\ref{FIG1X}(a)
shows total emitted power $P_\text{tot}$ as a function of wavelength
for various trench depths, and Fig.~\ref{FIG1X}(b) shows the
corresponding upwards ($P_\text{z+}$) and downwards ($P_\text{z-}$)
extracted powers.

Apparent in Fig.~\ref{FIG1X}(a) are a significant central wavelength
shift and a strong radiative rate modification. Indeed, as shown in
Fig.~\ref{FIG1X}(c), the central wavelength blue shifts more than 40
nm for depths increasing from 95 nm to 190 nm. The resonance full
width at half maximum (FWHM), correspondingly, decreases, indicating
an increase in field confinement and, consequently, the Purcell
Factor ($F_p$), Fig.~\ref{FIG1X}(e). The increased field confinement
for deeper trenches is a consequence of better overlap of the guided
field inside the slab and the etched region, which leads to
increased guided wave reflectivity and reduces coupling to
out-of-plane radiation.
Figures~\ref{FIG1X}(b) and (f) show the effect of grating asymmetry
on the ratio between upwards and downwards emitted powers. For a
symmetric grating with $d=t$, emission in both directions is the
same. Upwards emission is maximized for $d/t\approx0.8$, for which
$P_{z+}/P_{z-}=2.19$. This ratio increases to 5.8 for $d/t=0.7$,
however for a reduced $F_p$. Clearly, a trade-off must be reached
between Purcell enhancement and asymmetric emission.
\subsection{Effective mode volume calculation} The bullseye cavity's
effective mode volume $V_\text{eff}$=1.29$(\lambda_c/n)^3$ quoted in
the text was obtained with the expression
\begin{equation}
V_\text{eff} =
\frac{\int\epsilon(\mathbf{r})\left|\mathbf{E}(\mathbf{r})\right|^2d^3r}{\max\left\{\epsilon(\mathbf{r})\left|\mathbf{E(\mathbf{r})}\right|^2\right\}},
\label{veff_eq}
\end{equation}
where $\mathbf{E}$ is the modal electric field and
$\epsilon(\mathbf{r})$ the medium permittivity.  From
$V_{\text{eff}}$ and the calculated cavity $Q\approx200$, we can
estimate the maximum Purcell enhacement factor
$F_p$~\cite{ref:Gerard1}:
\begin{equation}
F_p=\frac{3Q(\lambda_c/n)^3}{4\pi^2V_\text{eff}}
\end{equation}
where $n$ is the refractive index and $\lambda_c$ is the cavity mode
wavelength.  The value determined through this calculation,
$F_p=11.8$, corresponds well with the value $F_p=11.0$ that we
determine from simulations of the enhanced radiated power for an
electric dipole in the bullseye.  The above values for $F_p$ assume
perfect dipole orientation with respect to the cavity field and
optimal dipole location within the field (i.e., in the bullseye
center).
\subsection{Fabrication} Based on simulation parameters, gratings
were fabricated on a $t$=190~nm thick GaAs layer containing a single
layer of self-assembled InAs quantum dots on top of a 1~$\mum$ thick
AlGaAs sacrificial layer. The epiwafer was grown with molecular beam
epitaxy, and displayed a quantum dot density gradient from
$>100~\mum^{-2}$ to 0$~\mum^{-2}$ along the $(01\bar{1})$ direction.
Electron-beam lithography was used to define the patterns, and a
single, timed, inductively-coupled plasma reactive ion etch
(ICP-RIE) step with an Ar/Cl$_2$ chemistry transferred the gratings
into the GaAs. This step was optimized so that the GaAs would be
partially etched to a desired depth in the grating region (see
Fig.1(b)), and fully etched over the large, curved rectangles just
outside the grating region seen in Fig.1(a). These open areas were
included to give access to the AlGaAs sacrificial layer, which was
etched with Hydrofluoric acid in a final step. Devices with varying
quantum dot (QD) densities were produced by fabricating the devices
along the QD density gradient of the wafer.

\subsection{Photoluminescence spectra and collection efficiency} The
spectra shown in Fig. 3 were obtained with a grating spectrometer
and a Si charge-coupled device (CCD). To obtain the PL intensity
from the excitonic lines $\text{X}_1$ and $\text{X}_2$ in Fig. 3(a)
without contributions from the broad cavity background, Lorentzians
were fitted to the corresponding peaks. Emitted photon rates plotted
in Fig. 4(a) for $\text{X}_1$ and $\text{X}_2$ and for unpatterned
GaAs QD lines correspond to the Lorentzian areas. The integrated
cavity photon rates in Fig. 4(b) were obtained by integrating the
spectra between 930 nm and 955 nm and subtracting rates from the two
excitonic lines.

To estimate the collection efficiency, we convert detected CCD
counts to photon counts into our NA=0.42 objective.  The two are
related by a conversion factor that is equal to the product of our
detection efficiency $\xi$, the transmission through the PL setup
$T_{\text{path}}$, and the transmission through the cryostat windows
$T_{\text{windows}}$. $\xi$ includes the in-coupling efficiency into
the spectrometer, the spectrometer's grating efficiency, and the
CCD's quantum efficiency, and is determined by sending a reference
laser with known power and wavelength into the spectrometer.  In
particular, we attenuate a 102 nW laser at 960 nm by 50 dB and
acquire a spectrum with a 1 s integration time. Integrating the
laser spectrum yielded a count rate of
$6.46\times10^4~\text{s}^{-1}$, which, when compared to the photon
rate just before the spectrometer ($4.93\times10^6~\text{s}^{-1}$),
gives a factor of $\approx 77$ photons per count.  Because the QD
emission wavelength ($\lambda_{\text{X}_1}=941$ nm) differs from the
calibration wavelength of 960 nm, we multiply this conversion factor
by 0.78, which is the ratio of the manufacturer-specified CCD
quantum efficiencies at these two wavelengths.  We therefore get an
overall detection efficiency $\xi=0.0167$ ($\equiv60$ photons per
count).

The transmission through the optical path (including the collection
objective) was $T_\text{path} = 0.156$, and was measured by
launching a fiber-coupled laser of known power and wavelength
through the optical setup. Note that, since the fiber NA=0.13 is
less than the objective's NA, all of the power emitted from the
fiber is collected, so the measured transmission includes only
transmission losses through the objective and routing optics.
$T_\text{windows}\approx0.87$, and includes transmission through the
radiation shield and outer cryostat windows.

The QD line $\text{X}_1$ in Fig. 4(a) yields a saturated CCD count
rate of
$R_{\text{X}_1}=1.14\times10^4~\text{s}^{-1}\pm0.16\times10^4~\text{s}^{-1}$,
where the uncertainty is a 95~\% fit confidence interval, due to
spectrometer resolution and detection noise. Assuming 100 \% QD
quantum efficiency, the rate of single photons emitted by the
saturated QD is equal to the pulsed pump excitation rate, like
$R_{\rm ex} = 50$~MHz. The collection efficiency was then calculated
as
\begin{equation}
\eta = \frac{R_{\text{X}_1}}{R_{\rm ex}\cdot \xi \cdot
T_\text{path}\cdot T_\text{windows}}, \label{eq2}
\end{equation}
Substituting all these values yields a collection efficiency
$\eta=10.1~\%\pm1.4~\%$.

\subsection{Quantum dot temperature dependence} The temperature
dependence of the sharp features $\text{X}_1$, $\text{X}_2$ and
$\text{X}_3$ in Figs. 3(a) and (b) was fitted with the Bose-Einstein
expression
\begin{equation}
\label{BE_eq}
E_{res}(T)=E_{res}(T=0)-S\hbar\omega\left[\coth\left(\frac{\hbar\omega}{2k_BT}\right)\right].
\end{equation}
This expression models the evolution of the semiconductor bandgap
energy with temperature due to electron-phonon interaction, assuming
no phonon dispersion, and has been successfully applied towards
excitonic transitions in epitaxially grown quantum
dots~\cite{ref:ortner_PhysRevB.72.085328,ref:kroner_temperature_2009}.
In Eq.~(\ref{BE_eq}), $T$ is the sample temperature, $E_{res}(T)$ is
the excitonic resonance energy, $\hbar\omega$ is the phonon energy,
and $S$ is a dimensionless coupling constant. The fits shown in
Fig.~3(c) were obtained with the parameters in Table~\ref{BE_table}.

The phonon energies $\hbar\omega$ in Table~\ref{BE_table} are
between the bulk GaAs transverse acoustic phonon energies at the X
and L points, $\hbar\omega_{TA}(\text{X})=7.7$~meV and
$\hbar\omega_{TA}(\text{X})=9.8$~meV, and the coupling coefficients
$S$ are compatible with those reported in refs.~
\onlinecite{ref:ortner_PhysRevB.72.085328} and
\onlinecite{ref:kroner_temperature_2009}. This indicates that the
sharp spectral lines correspond to excitonic QD transitions.

\begin{table*}[h]
\begin{center}
\begin{tabular}{c | c  c  c}
& $\text{X}_1$ & $\text{X}_2$ & $\text{X}_3$ \\
\hline
$E_{res}(T=0)$ (eV) & $1.3186\pm0.7\times10^{-4} $ & $1.3173\pm0.4\times10^{-4} $ & $1.3161\pm0.3\times10^{-4} $ \\
$\hbar\omega$ (meV) & $9.1601\pm0.8632$ & $9.1373\pm0.4890$ & $8.7596\pm0.4536$ \\
$S$& $0.9093\pm0.1005$ & $0.9276\pm0.0581$ & $0.8662\pm0.0581$ \\
\end{tabular}
\caption{\label{BE_table}Fitting parameters for temperature
dependence of QD lines $\text{X}_1$, $\text{X}_2$ and $\text{X}_3$
with eq.(\ref{BE_eq}). Errors are 95~\% fit confidence intervals
(two standard deviations).}
\end{center}
\end{table*}

\newpage
%\listoffigures

\end{document}